\documentclass[pdflatex,sn-nature]{sn-jnl}

\usepackage{graphicx}%
\usepackage{multirow}%
\usepackage{amsmath,amssymb,amsfonts}%
\usepackage{amsthm}%
\usepackage{mathrsfs}%
\usepackage[title]{appendix}%
\usepackage{xcolor}%
\usepackage{textcomp}%
\usepackage{manyfoot}%
\usepackage{booktabs}%
\usepackage{algorithm}%
\usepackage{algorithmicx}%
\usepackage{algpseudocode}%
\usepackage{listings}%
\usepackage{anyfontsize}%
\usepackage{bookmark}%
\usepackage{tabularx}
\usepackage{array}
\newcolumntype{Y}{>{\raggedright\arraybackslash}X}

\theoremstyle{thmstyleone}%
\theoremstyle{thmstyletwo}%

\theoremstyle{thmstylethree}%

\begin{document}

\title{Mobility and Contact Networks Shape Epidemic Outcomes: A Large-Scale Agent-Based Modeling Study}


\author*[1]{\fnm{Kuldip Singh} \sur{Atwal}}\email{katwal@gmu.edu}

\author[1]{\fnm{Emma} \sur{Von Hoene}}\email{evonhoen@gmu.edu}

\author[2]{\fnm{Hossein} \sur{Amiri}}\email{hossein.amiri@emory.edu}

\author[1]{\fnm{Dieter} \sur{Pfoser}}\email{dpfoser@gmu.edu}

\author[2]{\fnm{Andreas} \sur{Z\"ufle}}\email{azufle@emory.edu}

\author[1]{\fnm{Taylor} \sur{Anderson}}\email{tander6@gmu.edu}

\affil*[1]{\orgdiv{Geography and Geoinformation Science}, \orgname{George Mason University}, \orgaddress{\street{4400 University Drive}, \city{Fairfax}, \postcode{22030}, \state{VA}, \country{United States}}}

\affil[2]{\orgdiv{Department of Computer Science}, \orgname{Emory University}, \orgaddress{\street{201 Dowman Drive}, \city{Atlanta}, \postcode{30322}, \state{GA}, \country{United States}}}


\abstract{Human mobility plays a central role in shaping contact patterns that drive infectious disease transmission, yet mobility is often simplified in agent-based models (ABMs) due to data and computational constraints. The effects of these simplifications on model outputs are poorly understood. In this study, we systematically examined how alternative mobility assumptions influence emergent contact networks and epidemic dynamics within a large-scale ABM. Using a synthetic population of one million agents representing an urban environment, we implemented five mobility models varying along two dimensions: activity patterns (empirically derived vs. randomized) and destination choice mechanisms (empirical popularity, distance-based, or random). Holding disease parameters constant, we found that mobility assumptions alone produced substantially different contact network structures and epidemic trajectories, including differences in peak prevalence and shifts in outbreak timing. {\unboldmath $R_0$} varied fivefold across models, and randomized activities with popularity-weighted destinations produced a peak about 30\% higher and 4--5 days earlier than models with empirically derived activity patterns. Importantly, these differences could not be attributed to agents simply moving more or less overall since aggregate movement volumes were broadly comparable across models. Instead, the contrasting dynamics arose from how mobility generates contact opportunities: specifically, who meets whom, where, and how often. These results suggest that in models where mobility has not been carefully calibrated, simulated epidemic outcomes and evaluations of interventions may depend substantially on mobility assumptions. Our findings underscore the importance of mobility model calibration and validation, particularly in policy-facing applications.}


\maketitle

\section{Introduction}\label{sec_intro}


Modeling plays a crucial role in epidemiology by enabling the analysis and prediction of infectious disease dynamics through simulation of outbreak scenarios \cite{kermack_contribution_1997,aylett-bullock_june_2021}. Such models support a better understanding of the underlying processes that give rise to disease dynamics, informing public health responses, and evaluating intervention strategies \cite{Brauer2008, kerr2021covasim, iranzo2021epidemiological, smith2018agent}. There is a wide range of modeling approaches, ranging from compartmental models based on differential equations to network and agent-based models (ABMs) that represent heterogeneity in contacts and behavior \cite{RevModPhys.87.925, PhysRevX.1.011001,leventhal2015evolution, cuevas2020agent}. In this work, we focus on ABMs, which represent individuals as autonomous agents whose interactions can be governed by spatial, social, or behavioral rules \cite{carpenter2009design, arduin2017agent, hunter2018open}.

Infectious disease transmission is fundamentally driven by who contacts whom and how often \cite{kermack_contribution_1997, peirlinck_outbreak_2020, mossong_social_2008}. For a given pathogen, contact patterns shape interactions between susceptible and infectious individuals, subsequently generating new infections \cite{estrada_covid-19_2020}. Consequently, the structure and dynamics of contact networks are central determinants of simulated disease outcomes \cite{holme2016temporal}. In ABMs, contacts are often an outcome of assumptions about movement and activity, whether via explicit trajectories in space, movement between contexts (e.g., home, work, school), or simplified rules that implicitly generate who meets whom relationships. This makes mobility, broadly defined as how agents move through space and contexts over time, a key modeling choice that can shape the emergent contact network and disease outcomes.

Because human mobility is heterogeneous, scale-dependent, dynamic, and difficult to measure empirically, many ABMs rely on simplifying assumptions \cite{Box01121976, gonzalez2008understanding, alessandretti2020scales}. Ideally, agent mobility in these models would be behaviorally realistic, adaptive over time, and calibrated to empirical mobility data. In practice, however, this is difficult: detailed and behaviorally rich mobility representations are often less scalable, and fine-grained mobility data, especially at the individual level, are rarely available~\cite{kulkarni201920,mokbel2024mobility}. As a result, existing ABMs frequently adopt approximations such as restricted commuting patterns \cite{germann_mitigation_2006}, distance-based proxies for point-of-interest selection \cite{chang_modelling_2020}, or pre-specified social networks representing daily contacts in workplaces and other activities \cite{hinch_openabm-covid19agent-based_2021}. For example, large-scale synthetic population models such as EpiSimS \cite{mniszewski2013understanding} integrate census-derived populations with activity schedules from travel surveys, but still rely on simplifying assumptions about location choice and contact formation. Similarly, comparative and stochastic ABMs often simplify mobility using distance-based or randomized movement patterns to approximate spatial behavior and mixing \cite{chen2023stochastic, ajelli2010comparing, borkowski2009epidemic}. Some models further abstract movement as random walks to represent dynamic contact patterns \cite{chu2021random}.
These choices are often necessary, but they raise a methodological question: how sensitive are emergent contact networks and epidemic outcomes to the mobility assumptions used to generate interactions in ABMs?

In summary, the literature indicates that human movement and activity are closely linked to physical contact and proximity, and therefore to infectious disease transmission. However, there remains a limited understanding of how specific mobility assumptions in ABMs shape the emergent contact network and resulting disease dynamics. In this study, we designed and implemented five variations of agent mobility within an ABM of infectious disease spread to systematically explore how changes in these assumptions affect disease outcomes. Our models are designed to isolate the causal role of mobility by holding population initialization and infectious disease parameters constant. Such an approach enables a controlled comparison of how different mobility assumptions generate distinct emergent contact networks and, consequently, different epidemic trajectories. We evaluated these effects through changes in network structure and the resulting disease curves, with particular emphasis on shifts in outbreak timing and peak intensity.

\section{Results}\label{sec_results}

We simulated the spread of a generalized respiratory infectious disease within a synthetic population of one million agents as they move and interact in an urban environment representing Fairfax County, Virginia. We isolated the role of human mobility in the disease transmission process by maintaining identical disease parameters across five different models: NHTS Advan, NHTS Distance, NHTS Random, Random Places, and Random Advan. These models vary mobility assumptions along two dimensions: 1) activity patterns, which dictate the daily schedules of what people do and when, and 2) destination choices, which determine where they go. The activity patterns were either modeled using empirical trip chains from the National Household Travel Survey (NHTS) or were generated randomly. The probability that a point-of-interest (PoI) was chosen as a destination was determined using three methods: 1) using real-world foot traffic data from Advan Research, 2) as a function of the distance from the agent’s home census tract, and 3) uniformly at random. Model names indicate how these two dimensions of human mobility are paired across different mobility models. For example, NHTS Advan combines empirical schedules (NHTS data) with destination choice weighted by destination popularity in empirical mobility data (Advan). NHTS Distance combines empirical schedules with destination choice weighted by a distance-decay function based on distance from the agent’s home. NHTS Random combines empirical schedules with random destination choice. In contrast, Random Advan combines randomized schedules with destination choice weighted by destination popularity in empirical mobility data, whereas Random Places randomizes both schedules and destinations. By comparing these models, we identified how specific mobility assumptions drive epidemic growth. Further details about the implementation of the models are provided in the Methods Section~\ref{sec_methods}.

\paragraph{Mobility assumptions alter epidemic dynamics}

\begin{figure}[h!]
\centering
{\includegraphics[width=0.99\linewidth]{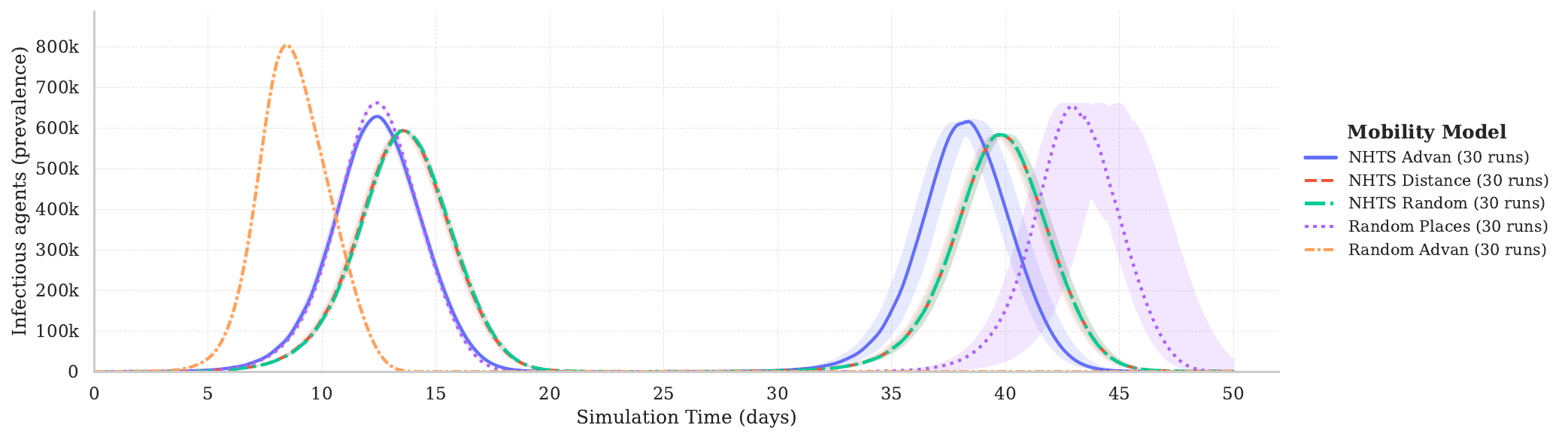}}
\caption{Disease outcome of all mobility models. Lines show the median number of infectious agents across 30 runs per model; shading shows the middle 95\% of runs.}
\label{disease_outcome}
\end{figure}

Random Advan peaked about 30\% higher and 4--5 days earlier than the NHTS-based models (Figure~\ref{disease_outcome}).
Comparison across the five mobility models showed that the different assumptions about human mobility substantially altered epidemic dynamics, affecting both the \textbf{peak prevalence} (the height of the peak on the y-axis) and the \textbf{time to peak} (the timing of the peak on the x-axis) of the infection. The NHTS-based models all produced reduced peak prevalence between 594--629k infectious agents with time to peak at approximately 12--14 days. The Random Advan and Random Places models yielded more acute peaks with peak prevalence at 804k and 662k and timing to peak occurring around days 8 and 12, respectively (Figure~\ref{disease_outcome}; Table~\ref{peak_metrics}). While all models except Random Advan exhibit secondary epidemic waves within the 50-day period (in Random Advan, infection died out after the first wave in all 30 runs), we note that our quantitative analysis throughout this section focuses on the first peak to isolate early transmission dynamics driven by mobility assumptions.

\paragraph{Epidemic metrics vary across mobility models}

To quantify the emerging disease dynamics in each model, we used the basic reproduction number $R_0$, epidemic growth rate, and doubling time. The basic reproduction number $R_0$ is defined as the average number of secondary cases generated during the simulation in a fully susceptible population by individuals who are infectious at the start of the simulation~\cite{lim2020interpretation}. We computed it as the mean number of new infections caused by each of the 100 initially infectious agents during its infectious period. The epidemic growth rate represents the per capita change in disease prevalence per unit time, providing a temporal measure of how rapidly an outbreak expands \cite{ma2014estimating}. This rate is typically expressed in units of day$^{-1}$, indicating continuous exponential growth rate on a daily time scale. Doubling time is the number of days required for the cumulative number of cases to double \cite{anzai2022doubling}. We estimated the growth rate $r$ by least-squares regression of the logarithm of the daily sum of infectious agents on day, up to the first-wave peak, and the doubling time as $\ln 2/r$. While $R_0$ captures the transmission potential of a pathogen, the growth rate and doubling time quantify the temporal velocity of an outbreak, which is crucial for guiding effective interventions \cite{pellis2021challenges}.

$R_0$ varied fivefold across mobility models, from 6.90 to 35.26. Table~\ref{disease_metrics} summarizes key epidemiological metrics across all mobility models, reporting the mean value and 95\% confidence intervals across 30 independent simulation runs. Random Advan exhibited the highest basic reproduction number $R_0$ (35.26) and growth rate of the first peak (1.17~day$^{-1}$), as well as the shortest doubling time (0.59~days), consistent with its early and pronounced epidemic peak. In contrast, the NHTS-based models displayed lower $R_0$ values (6.90--8.76), slower growth rate (0.70--0.76), and longer doubling times (0.91--0.99), resulting in wider and more flattened epidemic peaks. A doubling time shorter than one day corresponds to an exponential growth rate exceeding $\ln 2$ $\approx$  0.693 per day, consistent with the growth rates of 0.700--1.172 reported for all five models in Table~\ref{disease_metrics}.

\begin{table}[h]
\caption{Mean basic reproduction number ($R_0$), growth rate, and doubling time with 95\% confidence intervals across 30 independent simulation runs.}
\label{disease_metrics}
\centering
\begin{tabular}{lccc}
\toprule
\multirow{2}{*}{\textbf{Mobility Model}} & \textbf{Mean $R_0$} &  \textbf{Mean growth rate} & \textbf{Mean doubling time} \\
 &  & (first peak, day$^{-1}$) & (first peak, days) \\
\midrule
NHTS Advan & 8.76 [8.61, 8.91] & 0.758 [0.755, 0.760] & 0.91 [0.91, 0.92] \\
NHTS Distance & 7.01 [6.87, 7.14] & 0.700 [0.698, 0.701] & 0.99 [0.99, 0.99] \\
NHTS Random & 6.90 [6.76, 7.04] &  0.700 [0.698, 0.701] & 0.99 [0.99, 0.99] \\
Random Places & 11.31 [11.12, 11.49] & 0.768 [0.766, 0.770] & 0.90 [0.90, 0.90] \\
Random Advan & 35.26 [34.59, 35.94] &  1.172 [1.169, 1.174] & 0.59 [0.59, 0.59] \\
\bottomrule
\end{tabular}
\end{table}

\begin{table}[h]
\caption{Mean first-wave peak prevalence (infectious agents) and day of peak with 95\% confidence intervals across 30 independent simulation runs.}
\label{peak_metrics}
\centering
\begin{tabular}{lcc}
\toprule
\textbf{Mobility Model} & \textbf{Peak prevalence} & \textbf{Day of peak} \\
\midrule
NHTS Advan & 628,657 [628,343, 628,972] & 12.4 [12.4, 12.4] \\
NHTS Distance & 593,766 [593,482, 594,050] & 13.5 [13.5, 13.6] \\
NHTS Random & 594,135 [593,889, 594,380] & 13.6 [13.5, 13.6] \\
Random Places & 662,171 [662,014, 662,329] & 12.4 [12.4, 12.4] \\
Random Advan & 804,217 [803,529, 804,904] & 8.4 [8.4, 8.5] \\
\bottomrule
\end{tabular}
\end{table}

\paragraph{Mobility assumptions reshape contact network structure}
Across the five models, aggregate mobility characteristics remained broadly comparable, with total visits differing by at most 13\% (Table~\ref{mobility_metrics}). Total visit counts were similar across the NHTS-based and randomized models (approximately 160--180 million visits). The number of unique points of interest visited and the average dwell time were also comparable across all five models, although agents in the randomized models visited about two to three times as many distinct PoIs. Random Advan, the model producing the highest $R_0$ (35.26) and fastest epidemic growth, generated fewer total visits (approximately 160 million) than any of the NHTS-based models (approximately 180 million each), yet still produced the most intense outbreak. These results indicate that differences in epidemic outcomes cannot be attributed to agents simply moving more or less overall. Instead, the contrasting dynamics arise from how mobility generates contact opportunities, specifically, who meets whom, where, and how often,  rather than from differences in aggregate movement volume. To examine the mechanisms underlying these differences, we next analyzed the structural properties of the emergent contact networks.

\begin{table}[!h]
\centering
\footnotesize
\setlength{\tabcolsep}{4pt}
\caption{Mobility characteristics across the five mobility models during the simulation period. Total visits represent the cumulative number of visits  across all agents and time steps, counting repeated returns to the same location multiple times. Unique PoIs visited represent the number of different PoI locations visited at least once during the simulation. Average dwell time represents the mean duration of contiguous occupancy at a PoI, measured in hours. Trips per agent per day and distinct PoIs per agent (over the 50-day simulation) are means across agents.}
\label{mobility_metrics}
\begin{tabular}{lccccc}
\toprule
\multirow{2}{*}{\textbf{Mobility Model}} & \textbf{Total Visits} &  \textbf{Unique PoIs} & \textbf{Avg. dwell} & \textbf{Trips per} & \textbf{Distinct PoIs} \\
 & \textbf{(millions)} & \textbf{Visited} & \textbf{time (hours)} & \textbf{agent per day} & \textbf{per agent} \\
\midrule
NHTS Advan     & 179.98 & 325,907 & 6.67 & 3.58 & 49.4 \\
NHTS Distance  & 180.11 & 327,872 & 6.66 & 3.58 & 71.1 \\
NHTS Random    & 180.12 & 327,460 & 6.66 & 3.58 & 73.0 \\
Random Places  & 160.59 & 347,120 & 7.47 & 3.19 & 160.6 \\
Random Advan   & 159.99 & 346,667 & 7.50 & 3.18 & 132.3 \\
\bottomrule
\end{tabular}
\end{table}

Contact networks differed far more than mobility: mean degree varied more than fivefold across models (724.8--3873.8). Table~\ref{structural_metrics} summarizes the structural metrics of these networks across mobility models, reporting the mean value and 95\% confidence intervals across 30 independent simulation runs; for Random Advan, network metrics were computed from 5 of its 30 runs. Newman modularity quantifies the strength of a network's community structure by calculating the difference between the actual fraction of edges within groups and the expected fraction if edges were distributed at random \cite{newman2006modularity}. Average path length measures the mean shortest route between all node pairs, which can capture the structural resistance disease transmission must overcome to traverse a contact network. Average degree measures the mean number of distinct contacts per individual over the 50-day simulation and quantifies the interaction volume which can amplify $R_0$ and drive transmission capacity. Comprehensive reviews further explain how these structural properties influence epidemic dynamics \cite{RevModPhys.87.925, danon2011networks}.

\begin{table}[ht]
\centering
\footnotesize
\caption{Mean modularity, path length, and mean degree with 95\% confidence intervals across 30 independent simulation runs; for Random Advan, network metrics were computed from 5 of its 30 runs.}
\label{structural_metrics}
\begin{tabular}{lccc}
\toprule
\multirow{2}{*}{\textbf{Model}} & \textbf{Mean modularity } & {\textbf{Path length}} & \textbf{Mean degree} \\
 & \textbf{Score} (Q) & (avg.) & (avg.) \\
\midrule
NHTS Advan      & 0.5012 [0.5001, 0.5023]  & 2.854 [2.849, 2.859] & 923.1 [923.1, 923.1] \\
NHTS Distance   & 0.3371 [0.3355, 0.3386]  & 2.823 [2.819, 2.828] & 724.8 [724.8, 724.8] \\
NHTS Random     & 0.3373 [0.3358, 0.3388]  & 2.821 [2.817, 2.826] & 735.2 [735.2, 735.2] \\
Random Places   & 0.0323 [0.0322, 0.0324] & 2.481 [2.480, 2.481] & 856.9 [856.9, 856.9] \\
Random Advan    & 0.6179 [0.6178, 0.6181]  & 2.045 [2.044, 2.047] & 3873.8 [3873.5, 3874.0] \\
\bottomrule
\end{tabular}
\end{table}

Compared with the randomized models, the NHTS-based networks have longer average path lengths ($\approx$2.82--2.85 vs.\ 2.05--2.48) and slower epidemic growth ($\approx$0.70--0.76 vs.\ 0.77--1.17), consistent with studies showing that network structure can slow epidemic spread \cite{salathe2010dynamics}. The modularity of the NHTS-based models, well above that of Random Places, is likely due to their constrained schedules, which simulate recurring daily trajectories between the same home, work, and school locations and partition visits to places into distinct temporal windows, some lasting several hours. These schedules likely limit opportunities to mix with new contacts and instead produce stable communities of repeated interaction. 

\begin{figure}[H]
\centering
{\includegraphics[width=0.99\linewidth]{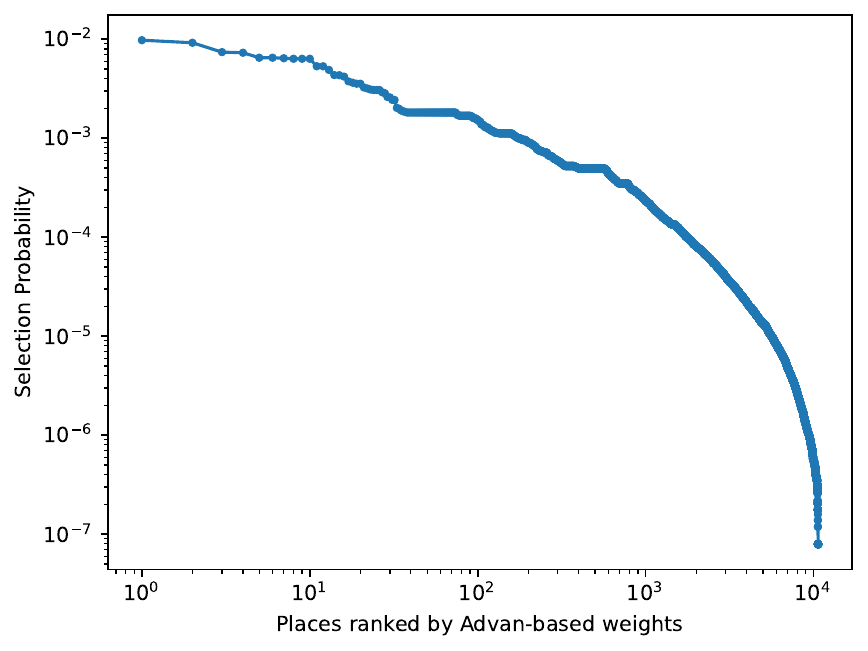}}
\caption{Selection distribution of destination places based on Advan weights: each place’s probability of being chosen, averaged over all Fairfax County home census tracts, with the five Advan-weighted place types (errands, restaurants, hospitals, recreation and schools) weighted equally.}
\label{visitors_dist}
\end{figure}

The strength and type of destination preference further differentiate these networks. NHTS Advan, in which destinations are weighted by their observed popularity in foot-traffic data, produces the highest and earliest epidemic peak among the NHTS-based models (629k on day 12.4). Advan weighting concentrates destination choice: the top 10\% of places carry 75\% of the selection probability, and the top 1\% carry 28\%, against 10\% and 1\% if every place were equally likely (Figure~\ref{visitors_dist}). Both Advan-based models choose public places and schools from these weights, and this concentration may contribute to NHTS Advan’s higher modularity (0.50) and mean degree (923) relative to the other NHTS models \cite{sun2013understanding,pathania2025leveraging}.

\begin{figure}[h!]
\centering
{\includegraphics[width=0.99\linewidth]{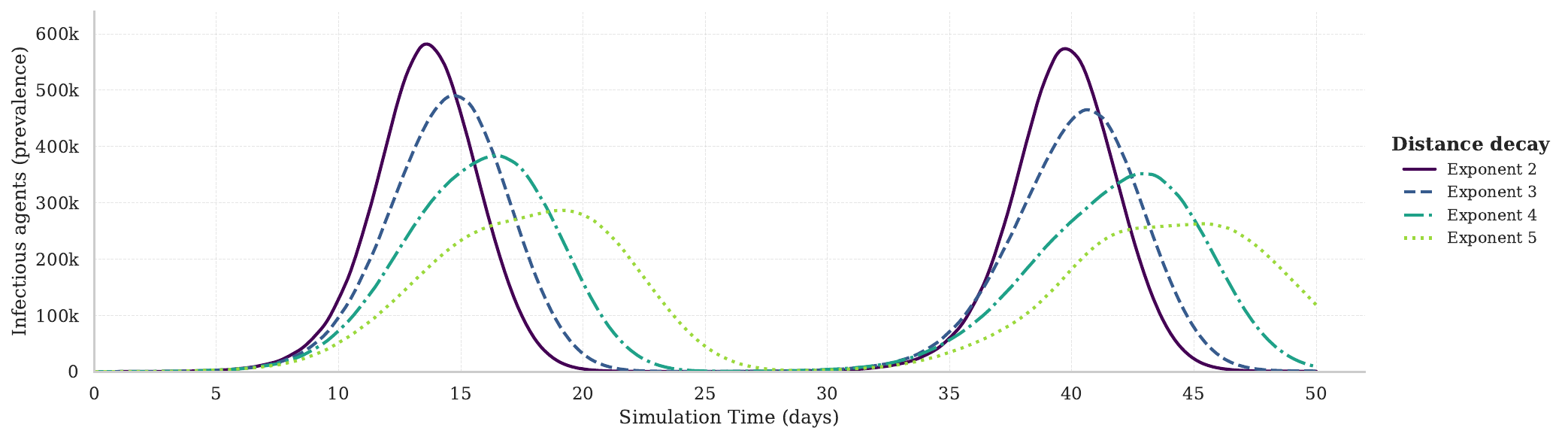}}
\caption{Disease outcome by varying power-law distance-decay in the NHTS Distance model (exponents 2--5; one run per exponent).}
\label{exp_decay_disease_outcome}
\end{figure}

NHTS Distance, in which destinations are weighted by a distance-decay function from the agent’s home, has a modularity (0.34) that is similar to NHTS Random (0.34). Their networks are nearly identical on every measure in Table~\ref{structural_metrics}, which likely explains why their epidemic curves are so similar. However, we note that the NHTS Distance result depends on the arbitrarily chosen distance decay. Increasing the exponent places a much greater preference for nearby destinations, increases modularity, and flattens the epidemic curve (see Figure~\ref{exp_decay_disease_outcome}; modularity 0.42 at exponent 2 to 0.81 at exponent 5).

Although the Random Advan contact network has the highest modularity of all five models ($Q = 0.62$), it also has the shortest average path length ($\approx$2.05) and an average degree that is more than 4 times higher than the other models ($\approx$3874). This high density likely emerges because individuals follow random activity patterns instead of structured daily schedules, while still concentrating visits at popular, high-traffic locations. This combination increases the number of distinct contacts and weakens the extent to which community boundaries constrain transmission. Consequently, the network supports rapid disease spread throughout the population, producing the highest estimated $R_0$ (35.26) and growth rate (1.17). These outcomes appear to be driven more by network density caused by unscheduled visits to popular places with a high concentration of agents than by modularity alone.

Finally, Random Places exhibits negligible modular structure ($Q \approx 0.03$), indicating near-homogeneous mixing across locations. This lack of community structure occurs because agents select a new destination uniformly at random for each trip. Unlike the NHTS-based models, in which individuals repeatedly return to the same home, workplace, and school assigned at the start of the simulation, this reshuffling increases the likelihood that individuals encounter new people during each trip. Although this reduces structural barriers to transmission, the resulting contact network remains sparser than that generated by Random Advan (average degree $\approx$857 vs.\ 3874). One reason may be that Random Places distributes agents more evenly across destinations, whereas Random Advan weights destination choice by popularity, concentrating agents at highly visited places and increasing contact opportunities. As a result, the Random Places network produces fewer opportunities for intense local mixing than Random Advan, leading to a lower estimated $R_0$ (11.31 vs.\ 35.26) and slower early epidemic growth (0.77 vs.\ 1.17).

\paragraph{Results Summary}

In summary, mobility assumptions in the ABM substantially shape the emergent contact network and, in turn, epidemic outcomes. 
Models based on structured empirical NHTS schedules produced networks with longer path lengths and slower epidemic growth than the randomized models, likely because recurring daily routines partition contact events into distinct temporal windows and limit co-location across the population. In contrast, Random Advan combined randomized schedules with popularity-weighted destinations to generate the densest network and the fastest epidemic growth. Together, these results suggest that destination popularity amplifies transmission most strongly when movement is not constrained by structured daily schedules.

\section{Methods}\label{sec_methods}

\subsection{Mobility Models Framework}

Human mobility encompasses multiple factors, including visit purpose, destination, timing, and duration of stay. Individuals typically follow structured routines, such as commuting to the same workplace each day, whereas other activities, including dining and errands, may vary depending on location popularity or proximity. To systematically understand how these mobility characteristics influence disease transmission, we designed five mobility models that vary along two key dimensions: activity patterns (i.e., what individuals do and when) and place selection mechanisms (i.e., where they go). The assumptions and the specific data sources for each model are described in Table~\ref{mobility_models}. These models range from empirical data–driven mobility to highly randomized mobility, exploring how different mobility assumptions lead to the formation of contact networks that ultimately affect the disease dynamics. The five models reflect real mobility assumptions commonly used in large-scale agent-based modeling frameworks \cite{mniszewski2013understanding, bisset2009epifast, ajelli2010comparing}. While we implement the models (see Section \ref{sec:implementation}) using the 2017 National Household Travel Survey (NHTS) \cite{nhts2017} for the empirical activity patterns and timing and the 2019 Advan Weekly Patterns \cite{AdvanResearch2026} and the 2015 LEHD Origin-Destination Employment Statistics (LODES) \cite{lodes2015} origin-destination data for the empirical place selection, we note that the framework could be implemented with alternative travel survey or origin-destination datasets.

\begin{table}[h]
\caption{Mobility models used in this study, specifying the conceptual basis for activity patterns, place selection, timing, and work/school assignment, with the data sources used in this implementation shown in parentheses.}
\label{mobility_models}
\centering
\footnotesize
\setlength{\tabcolsep}{4pt}
\renewcommand{\arraystretch}{1.2}
\begin{tabularx}{\textwidth}{l Y Y Y l}
\toprule
\textbf{Mobility Model} & \textbf{Activity Pattern} & \textbf{Place Selection} & \textbf{Timing} & \textbf{Work/School Assignment} \\
\midrule
NHTS Advan & Travel survey (NHTS) & Weighted by O-D flows (Advan) & Travel survey (NHTS) & Yes (LODES/Advan) \\
NHTS Distance & Travel survey (NHTS) & Distance-weighted from home & Travel survey (NHTS) & Yes (distance decay) \\
NHTS Random & Travel survey (NHTS) & Uniform selection & Travel survey (NHTS) & Yes (random) \\
Random Places & No activity structure & Uniform selection & Uniform distribution & No \\
Random Advan & Random activity & Weighted by O-D flows (Advan) & Uniform distribution & No \\
\bottomrule
\end{tabularx}
\end{table}

\begin{enumerate}
    \item \textbf{NHTS Advan:} The NHTS Advan model represents the most empirically grounded mobility. Agents followed daily activity schedules derived from a travel survey, including trip chains, start times, and dwell durations. The destination locations for activities were selected using empirically observed origin–destination (OD) mobility flow data, which provided weighted probabilities for these choices. Agents were assigned fixed homes, workplaces, and schools, to which they repeatedly return, reflecting daily routines. The destinations for other activities were selected at each occurrence using weighted probabilities from the OD flow data. This approach is closely aligned with large-scale synthetic population models such as EpiSimS framework, which integrates detailed activity schedules with data-driven location assignment to generate realistic contact networks \cite{mniszewski2013understanding}. By preserving both temporal schedules and spatial heterogeneity, the NHTS Advan model captures repeated interactions within structured environments alongside concentrated mixing at highly popular locations.   
    
    \item \textbf{NHTS Distance:} The NHTS Distance model retained the same empirically derived activity schedules but replaced data-driven destination selection with a spatial proximity rule. In this case, agents preferentially selected destinations based on a distance-decay function from their home census tracts. This reflects the tendency for individuals to favor nearby places. Similar to NHTS Advan, agents repeatedly returned to home, work, and school locations, while selecting other locations at each occurrence using distance decay. This formulation is similar to spatially explicit epidemic models that incorporate distance-dependent interaction methods or gravity-like mobility assumptions, where transmission intensity declines with geographic separation \cite{ajelli2010comparing, hoertel2020stochastic}. By isolating the role of spatial proximity while preserving temporal schedules, this model enables analysis of how localized movement patterns shape contact networks and epidemic spread.  
    \item \textbf{NHTS Random:} In the NHTS Random model, agents followed empirical daily schedules from the travel survey, but destination choices were uniformly randomized for each activity. Agents repeatedly return to homes, workplaces, and schools that were randomly assigned at the start of the simulation. Locations associated with discretionary activities, such as restaurants, recreation, errands, and healthcare, were randomly selected at each occurrence. Similar approaches have been explored in stochastic, schedule-based agent frameworks such as the ``scheduled walker'' model, which incorporates structured timing alongside flexible or randomized spatial behavior to examine how different mobility assumptions influence epidemic dynamics \cite{borkowski2009epidemic}.   
    
    \item \textbf{Random Places:} The Random Places model represents a fully unstructured mobility model in which agents did not follow any activity patterns and selected destination locations weighted uniformly. The dwell time of each destination was also sampled from a uniform distribution. Unlike the above models, agents did not return to fixed locations. This produces highly dynamic and continually reshuffled contact patterns, approximating homogeneous mixing over time. Such dynamics are aligned with random-walk-based epidemic models, in which agents move stochastically through space and form contacts via co-location, leading to rapidly mixing populations \cite{chu2021random}.   
    
    \item \textbf{Random Advan:} The Random Advan model combines randomized activity patterns with empirically grounded destination preferences. Agents selected activities and timing from random uniform distributions but chose public places and schools according to empirically derived weights from the OD mobility flows. This hybrid configuration captures the interaction between stochastic movement and heterogeneous location preferences, which may lead to concentrated mixing at high-traffic venues despite the absence of structured schedules. This approach is closely related to recent mobility network models that use fine-grained point-of-interest data to represent human movement and demonstrate how a small number of highly visited locations can disproportionately drive transmission dynamics \cite{chang2021mobility}.  
\end{enumerate}

\subsection{Contact and Disease Model}

Since the scope of this study is to understand the effects of mobility assumptions on disease modeling, we used a fixed set of parameters across all models that represent a generalized respiratory infectious disease (Table~\ref{disease_params}). Disease spread was modeled using a stochastic, agent-based susceptible–exposed–infectious–recovered–susceptible (SEIRS) framework, in which transmission occurs through co-location of agents at shared locations. Co-location is defined as agents being present at the same PoI during a given time tick. At each time step, co-located agents at the same place (e.g., households, workplaces, or public venues) form contact groups at each location. Within these groups, infectious agents may transmit the disease to susceptible individuals.

\begin{table}[h]
\caption{Fixed disease parameters used across all simulation models.}
\label{disease_params}
\centering
\begin{tabular}{l l}
\toprule
\textbf{Parameter} & \textbf{Value} \\
\midrule
Number of initially infectious agents & 100 \\
Probability of infection per contact per hour & 0.10 \\
Exposed state duration & Uniform distribution between 1--2 days \\
Infectious state duration & Uniform distribution between 1--5 days \\
Immunity duration before returning to susceptible & Uniform distribution between 10--20 days \\
\bottomrule
\end{tabular}
\end{table}

To capture realistic limits on interactions in crowded environments, we constrained contact events per agent to a sublinear function of local population size: if $n$ agents are co-located, each interacts with at most $\log_2(n)$ other agents (rounded to the nearest integer). This reflects bounded contact rates in large gatherings, where individuals engage with a limited subset rather than all nearby agents, consistent with prior work on saturating interaction networks \cite{zhang2003global, zhang2015modeling, saadi2025agent}. At each simulation tick, every infectious agent interacts with the subset of co-located agents chosen via a fixed ordering set at initialization; this subset remains constant while agents stay co-located, modeling stable, repeated interactions (such as in workplaces) that increase repeated exposure. Each infectious agent draws once per one-hour tick and, with probability 0.10, infects all of its susceptible contacts in that tick.

The disease progression follows four states: susceptible, exposed, infectious, and recovered. Susceptible agents become exposed upon effective contact with infectious agents. The exposed (latent) period is drawn from a uniform distribution between 1 and 2 days, after which agents become infectious. The infectious period is uniformly distributed between 1 and 5 days, during which agents can transmit the disease. Afterward, agents enter the recovered state and acquire temporary immunity lasting between 10 and 20 days (uniformly distributed). Once immunity wanes, agents return to the susceptible state and may be reinfected.

\subsection{Model Implementation}
\label{sec:implementation}

\paragraph{Population}
We synthesized a population of one million agents representing Fairfax County, Virginia, US, including 480,000 workers, 210,000 students, and 310,000 homemakers. The agent population was generated using iterative proportional fitting (IPF) population synthesis. This approach fits respondents in the 2017 National Household Travel Survey (NHTS) \cite{nhts2017} based on their demographic characteristics (age, race, income, education, sex, student status, worker status) to the marginal totals for those same characteristics observed in each of Fairfax County's census tracts. The trip characteristics of each individual are carried over into the final synthetic population, to be used by the models that leverage the NHTS trip chains for activity selection. 

\paragraph{Environment}

The simulation environment consisted of 347,120 PoI locations spanning residential, workplace, and public venues. These include over 318,000 homes, approximately 17,800 workplaces from the official building footprint data \cite{ffxofficial}. This data was aggregated into the type ``residential'' from single-family and multi-family homes. The type ``workplace'' was aggregated from commercial, public, industrial, and other types in the raw data. Centroids of the building footprints were used as PoIs and were combined with a set of public places derived from Advan PoI data, such as restaurants, recreational venues, hospitals, and errand-related facilities. Related facilities in the Advan PoIs were aggregated into broader categories. For example, fitness centers, museums, parks, and religious institutions are grouped under recreation, and doctors' offices and hospitals are combined into hospitals. This combined dataset was used as the physical infrastructure for agents to visit places (Table~\ref{places_combined}).

\begin{table}[h]
\caption{Aggregated place categories used in the simulation, mapping NHTS trip purposes to PoI types unified from building footprint data and Advan PoIs. NHTS trip purposes define activity types, while PoI types represent the corresponding locations where activities occur. Counts represent the final number of PoIs after integrating all data sources. NAICS codes are provided where applicable.}

\label{places_combined}
\centering
\footnotesize
\begin{tabular}{l l r l}
\toprule
\textbf{NHTS Trip Purpose} & \textbf{PoI Type} & \textbf{Count} & \textbf{Source Categories} \\
(Code) & & &  (NAICS Prefix) \\
\midrule
Home (01) & Residential & 318,201 & Single-family, Multi-family\\
& & & residential buildings \\
Work (10) & Workplaces & 17,803 & Commercial, Public, Industrial,\\
& & & Other buildings \\
School/Daycare (20) & Schools & 1,161 & Schools (61), Childcare (6244) \\
Meals (80) & Restaurants & 2,165 & Restaurants (7225) \\
Social/Recreational (50) & Recreation & 1,621 & Religious (8131), Museums \& parks (712), \\
& & & Fitness (713)\\
Medical/Dental services (30) & Hospitals & 1,265 & Doctors’ offices (621), Hospitals (622) \\
Shopping/Errands (40), Other (97) & Errands & 4,904 & Retail and service establishments (445) \\
\midrule
\textbf{Total} & & \textbf{347,120} & \\
\bottomrule
\end{tabular}
\end{table}

All models except Random Places used types and locations of the PoIs to go from one place to another. For example, in NHTS Advan, an agent selects and goes to a restaurant PoI if the activity chain lists ``restaurants’’ as the next activity. In Random Places, agents only select a PoI location for mobility without considering the type of PoIs. We simulated activities of the agents according to the NHTS trip purposes \cite{nhtscodes}, which were mapped to the PoI types in the unified input dataset of PoIs (Table~\ref{places_combined}). 

\paragraph{Home, School, and Workplace Assignment}
The population synthesis approach assigns each agent to a census tract until the synthetic population count matches that of the real population. Then, agents were randomly assigned to residential buildings within their home census tract, with at most four agents per building. While all five models start with the same population and home assignment, only agents in the NHTS models are assigned work and school locations that they repeatedly visit. 

In NHTS Advan, worker agents were assigned fixed workplaces based on the 2015 LEHD Origin-Destination Employment Statistics (LODES) O-D commuting data \cite{lodes2015}. Commuting flows aggregated at the census tract level were used to derive probabilities of travel between home and work tracts, after which specific workplace locations were randomly selected within destination tracts. Schools were assigned to student agents based on weights derived from Advan weekly patterns data from October 21--27, 2019 \cite{AdvanResearch2026}, computed by normalizing visitor counts from origin census tracts to each PoI that week. Visitor counts shared by several PoIs, such as stores in one mall, were split equally among them. For the NHTS Distance model, work and school locations were assigned using a distance-decay function from their home census tracts, normalized so that the probabilities for each home census tract add up to 1. Figure~\ref{exp_decay_disease_outcome} uses power-law decay, $P(d) \propto d^{-\beta}$, with $\beta \in \{2,3,4,5\}$. For the NHTS Random model, these locations were assigned at random and remained fixed throughout the simulation.

\paragraph{Activity Pattern and Timing}
In NHTS Advan, NHTS Distance, and NHTS Random models, agents repeatedly follow a daily activity schedule using empirical trip chains, start times, and dwell times from the 2017 National Household Travel Survey (NHTS) \cite{nhts2017}. This activity schedule is assigned to the agents during the population synthesis stage. Each day, each agent participates in up to seven activity types, including home, work, school, restaurants, recreation, hospitals, and errands.

For the NHTS-based models, start times and dwell durations (reported in the NHTS survey in minutes) were converted to hourly intervals to match the simulation tick time. To prevent short-duration activities from being eliminated, dwell times were rounded to the nearest hour with an enforced minimum of one hour. For example, all durations between 1 and 89 minutes were assigned 1 hour; 90 to 149 minutes were assigned 2 hours. Daily schedules were constructed sequentially by assigning the first activity of the day to its rounded start time. Subsequent activities followed immediately after the preceding activity’s discretized duration elapsed, without explicit modeling of travel time between activities. The resulting daily activity schedules were mapped onto a 24-hour cycle and repeated identically across simulation days.

 In Random Places, agents do not select any activities while in Random Advan, agents randomly select activities for which they dwell before randomly selecting another. In these models, agents independently sample their start time and dwell durations from uniform distributions bounded between four and ten hours. The bounds were calibrated using the average number of trips an agent makes during the simulation period in the NHTS-based models. After the dwell duration elapses, an agent randomly selects an activity to go to the next place.

\paragraph{Place Selection}
In NHTS Advan and Random Advan, we leveraged the weights derived from the Advan foot traffic data for destination choices at public places (restaurants, recreation, errands and hospitals) and schools \cite{AdvanResearch2026}. In NHTS Distance, agents selected destinations based on a distance-decay function from their home census tracts. In NHTS Random, destinations were selected at random, but corresponding to the activity type. In Random Places, agents randomly selected destination places for each trip, while in Random Advan, public places and schools were selected based on Advan weights.

\paragraph{Initialization}

All models included a synthetic population of one million agents. Initially, 100 agents within a single census tract were randomly selected to be infectious. We ran the models for a 50-day simulation period, where each simulation tick represents one hour of real-world time. This allows daily activity patterns and interactions to be modeled at an hourly temporal resolution. All mobility models were implemented using the Repast4Py agent-based modeling platform, enabling scalable simulation of large synthetic populations and their interactions \cite{collier2022distributed}. Contacts observed during the 50-day simulation were used to construct simple, undirected graphs in which nodes are agents and an edge joins two agents if either selected the other as a contact at least once; these graphs were analyzed using the Python iGraph library \cite{csardi2006igraph}. Communities were found with the Louvain method using a fixed seed \cite{Blondel_2008}, and average path length was estimated from sampled source nodes. Simulation results were averaged over 30 stochastic runs per model, except for Random Advan’s contact network metrics (computed from 5 of its 30 runs) and mobility metrics (12 of its 30 runs), which were limited by the memory needed to process its contact networks. Each run used a distinct base seed, from which every process derived its own seed; only the initial cases were selected with a fixed seed, so every run starts from the same 100 infectious agents.

\section{Discussion and Conclusion}\label{sec_discussion}

This study demonstrates that assumptions about human mobility in ABMs can determine the emergent contact network structure that affects epidemic dynamics. By systematically varying activity schedules and destination choice mechanisms, while holding disease parameters constant, we isolate how mobility representations can produce substantially different epidemic trajectories. These findings extend prior work by explicitly linking mobility, network structure, and disease outcomes within a single framework. A key takeaway is that, with disease parameters held fixed, mobility representation alone substantially changed epidemic outcomes in these models, a dependency that is easy to overlook when mobility assumptions are treated as background choices rather than consequential modeling decisions.

A key implication of our results is that activity schedules in human mobility can shape network structure, which in turn may slow epidemic spread. Models grounded in NHTS-based schedules consistently produced networks with longer path lengths than the randomized models, leading to lower epidemic peaks. This aligns with findings from \cite{block_social_2020}, where longer path length in social networks reduces transmission efficiency and flattens epidemic curves. In contrast, removing temporal regularity combined with heterogeneous destination preferences increases network connectivity and accelerates transmission. The Random Advan model illustrates how the interaction between random activity patterns and popularity-driven destination choice can produce highly connected networks. This result complements mobility-driven ABMs such as \cite{muller_realistic_2020}, which incorporate empirical movement data to reproduce real-world epidemic dynamics.

An important result is observed in the Random Places model, which produces a higher peak prevalence than NHTS Advan (662k vs.\ 629k) despite fewer distinct contacts per person (mean degree 857 vs.\ 923). This highlights that epidemic outcomes are not determined solely by the number of distinct contacts, but also by how contact events are distributed across time and individuals. Fully randomized mobility leads to near-homogeneous mixing over time, increasing the likelihood that infections spread broadly across the population. This observation reinforces insights from network-based studies such as \cite{small_modelling_2020}, which emphasize that changes in contact patterns, not just contact frequency, can significantly alter epidemic behavior. 

By decomposing mobility into its key components, activity timing and destination selection, and showing how each independently and jointly shapes contact network structure, our framework offers a clearer explanation for why different ABMs with different mobility assumptions may yield divergent epidemic predictions \cite{aleta_modelling_2020, salem_multi-agent-based_2022}.

These results underscore the need for careful consideration of mobility representations and model calibration, particularly in policy-facing models where predictions inform public health decisions. Crucially, because the differences we observe arise from structural properties of contact networks rather than from aggregate movement volume, they are less detectable through standard model evaluations that focus on disease parameters alone. This suggests that mobility assumptions may represent a hidden source of uncertainty in many existing ABMs, one that is not routinely reported or tested, a concern amplified by evidence that, among ABMs developed to model COVID-19, a large fraction were neither calibrated nor validated \cite{von2026all}.

These findings carry a cautionary implication for the use of ABMs in practice. Where mobility assumptions have not been calibrated or validated against empirical data, simulated epidemic outcomes may be driven substantially by those assumptions. This is particularly consequential for intervention evaluation:  if the contact network generated by the model does not reflect real-world interaction patterns, then an intervention that appears effective in simulation may not be in practice.
While our study cannot determine which mobility model best approximates real-world behavior, since that would require validation against observed outbreak data, the range of outcomes we observe across models suggests that this uncertainty is large enough to affect policy-relevant conclusions. We therefore advocate for explicit reporting and sensitivity testing of mobility assumptions in ABMs, especially when model outputs are used to inform public health decisions.

This study has several limitations, many of which arise from the design choices to isolate the effects of mobility assumptions on epidemic dynamics, rather than to reproduce real-world mobility or disease outcomes. First, the mobility models consider a simplified set of seven activity categories and do not represent transportation or additional activity types, potentially omitting mobility pathways that contribute to contact formation. Second, simulations are conducted on a uniform daily cycle without distinguishing between weekdays and weekends, thereby not capturing temporal variations in routine behavior. Third, the input datasets may contain sampling biases, particularly in passively collected mobility data; however, they are primarily used to construct relative mobility patterns rather than to reproduce exact population-level behavior. Fourth, the disease model represents a generalized infectious outbreak with fixed parameters and is not calibrated to a specific pathogen; therefore, the resulting epidemic dynamics are not intended to match real-world outbreaks and would likely differ under disease-specific parameterization. Finally, all experiments are conducted within the spatial boundaries of Fairfax County, and the findings may not generalize to regions with different demographic, geographic, or mobility characteristics. Where official building or commuting data are unavailable, building types and commuting flows predicted from OpenStreetMap could support applying the framework elsewhere \cite{atwal2022predicting,atwal2025commuting}. Nonetheless, the framework is intended to provide general insights into how mobility representations likely shape contact networks and epidemic outcomes, rather than location-specific predictions.

Overall, this study highlights the broader insight that epidemic dynamics emerge from the interplay between human mobility behavior and network structure, rather than from disease parameters alone. While prior work \cite{estrada_covid-19_2020} has emphasized the integration of compartmental and network approaches, our results suggest that accurately modeling the processes, such as human mobility, that generate contact networks is essential for capturing realistic disease dynamics. More broadly, our findings suggest that the field would benefit from treating mobility representations as a first-class modeling choice, subject to the same scrutiny as disease parameters, including explicit reporting of assumptions, sensitivity testing across alternative representations, and wherever possible, validation against empirical contact or outbreak data. Future research should extend this framework by incorporating additional behavioral factors, such as adaptive responses to risk, demographic heterogeneity, and transportation systems, as well as by validating mobility assumptions against real-world disease data, a step that would move this line of work from sensitivity analysis to ground-truth evaluation.

\backmatter


\bmhead{Acknowledgements}

This project was supported by resources provided by the Office of Research Computing at George Mason University (URL: \url{https://orc.gmu.edu}). We thank Dr. Sandro M. Reia for insightful discussions.

\section*{Declarations}

\begin{itemize}


\item The code is available in a GitHub repository at \url{https://github.com/heykuldip/mobility-affects-disease} and archived at \url{https://doi.org/10.5281/zenodo.23407174}. The container image used for the simulations is archived at \url{https://doi.org/10.5281/zenodo.23406362}.


\end{itemize}


\end{document}